\documentclass[10pt, conference, a4paper]{IEEEtran}
\IEEEoverridecommandlockouts

\usepackage[utf8]{inputenc} 
\usepackage[T1]{fontenc}
\usepackage{url}
\usepackage{ifthen}
\usepackage{cite}
\usepackage[cmex10]{amsmath} %
\usepackage{mleftright}
\usepackage{bbm}
\usepackage{amssymb, amsfonts}
\usepackage{dsfont}
\usepackage{nicefrac}
\usepackage{aligned-overset}
\usepackage{graphicx}         %
\usepackage{caption}
\usepackage{makecell}

\usepackage[nolist]{acronym} 
\usepackage[normalem]{ulem}
\usepackage{pifont}
\usepackage{nicematrix}
\usepackage{booktabs}
\usepackage{tabularx}
\usepackage{multirow}
\usepackage{array}

\usepackage{tikz}
\usetikzlibrary{arrows,calc,shapes,graphs,graphs.standard,quotes,arrows.meta,decorations.markings,positioning,fit,positioning,hobby,decorations.text,decorations.pathreplacing,dsp}

\usepackage{pgfplots}
\pgfplotsset{compat=1.18}
\pgfdeclarelayer{background}
\pgfdeclarelayer{foreground}
\pgfsetlayers{background,main,foreground}
\usepgfplotslibrary{groupplots}
\tikzset{>=latex}

\pgfplotsset{
/pgfplots/short line/.style={
    legend image code/.code={
        \draw[mark repeat=2,mark phase=2,#1] 
            plot coordinates {(0cm,0cm) (0.1435cm,0cm) (0.287cm,0cm)};
    },
},
}

\usepackage{setspace}
\usepackage{algpseudocode}
\usepackage[linesnumbered,titlenumbered,ruled,vlined,resetcount]{algorithm2e}

\usepackage{etoolbox}
\makeatletter
\patchcmd\algocf@Vline{\vrule}{\vrule \kern-0.4pt}{}{}
\patchcmd\algocf@Vsline{\vrule}{\vrule \kern-0.4pt}{}{}
\makeatother

\IncMargin{0.125em}

\makeatletter
\patchcmd{\@algocf@start}{-1.5em}{-1.25em}{}{} %
\makeatother

\newcommand{\tikzgets}{\tikz[baseline=-0.5ex] \draw[-{Straight Barb[scale=0.8]}, yshift=0.05ex] (0.3,0) -- (0,0);}
\renewcommand{\gets}{\mathrel{\tikzgets}}

\newcommand{\tikzto}{\tikz[baseline=-0.5ex] \draw[-{Straight Barb[scale=0.8]}, yshift=0.05ex] (0.0,0) -- (0.3,0);}
\renewcommand{\to}{\mathrel{\tikzto}}

\SetKwProg{Fn}{subroutine}{:}{end}
\let\oldnl\nl%
\newcommand{\nonl}{\renewcommand{\nl}{\let\nl\oldnl}}%
\newcommand{\algrule}[1][.2pt]{\par\vskip.5\baselineskip\hrule height #1\par\vskip.5\baselineskip}

\usepackage{amsthm}
\newtheorem{corollary}{Corollary}%
\newtheorem{theorem}{Theorem}%
\newtheorem{lemma}{Lemma}%
\newtheorem{definition}{Definition}%

\usepackage{marginnote}

\SetAlCapNameFnt{\footnotesize}
\SetAlCapFnt{\footnotesize}

\begin{acronym}
\acro{ABLER}{asymptotic BLER}
\acro{BCH}{Bose--Ray-Chaudhuri--Hocquenghem}
\acro{BER}{bit error rate}
\acro{BI-AWGN}{binary-input additive white Gaussian noise}
\acro{BLER}{block error rate}
\acro{BPSK}{binary phase shift keying}
\acro{CDF}{cumulative distribution function}
\acro{CW}{Cordaro--Wagner}
\acro{DE}{density evolution}
\acro{DEGA}{DE Gaussian approximation}
\acro{FHT}{fast Hadamard transform}
\acro{LLR}{log-likelihood ratio}
\acro{LSB}{least significant bit}
\acro{ML}{maximum likelihood}
\acro{NPRS}{naturally punctured repeated simplex}
\acro{NPRSD}{NPRS dual}
\acro{OSD}{ordered statistic decoding}
\acro{ORCAS}{optimally recursively concatenated simplex}
\acro{PPV}{Polyanskyi--Poor--Verdú}
\acro{RM}{Reed--Muller}
\acro{SC}{successive cancellation}
\acro{SSC}{simplified SC}
\acro{SNR}{signal-to-noise-ratio}
\acro{SPC}{single parity check}
\end{acronym}

\definecolor{mittelblau}{RGB}{0, 126, 198}
\definecolor{violettblau}{cmyk}{0.9, 0.6, 0, 0}
\definecolor{rot}{RGB}{238, 28 35}
\definecolor{apfelgruen}{RGB}{140, 198, 62}
\definecolor{gelb}{RGB}{255, 229, 0}
\definecolor{orange}{RGB}{244, 111, 33}
\definecolor{pink}{RGB}{237, 0, 140}
\definecolor{lila}{RGB}{128, 10, 145}
\definecolor{hellgrau}{RGB}{224, 224, 224}
\definecolor{mittelgrau}{RGB}{128, 128, 128}
\definecolor{dunkelgrau}{RGB}{80,80,80}
\definecolor{anthrazit}{RGB}{19, 31, 31}
\definecolor{aqua}{RGB}{0, 255, 255}

\renewcommand{\mid}{\,\vert\,}
\newcommand{\tp}{^{\mathsf{T}}}
\newcommand{\lb}{l_\mathrm{b}}
\newcommand{\w}{w_\mathrm{H}}

\begin{document}
\setlength{\columnsep}{0.201in} %

\title{ORCAS Codes: A Flexible Generalization of\\Polar Codes with Low-Complexity Decoding}

\author{\IEEEauthorblockN{Andreas Zunker, Marvin R\"ubenacke, and Stephan ten Brink}
\IEEEauthorblockA{Institute of Telecommunications, 
University of Stuttgart, Germany\\
\{zunker, ruebenacke, tenbrink\}@inue.uni-stuttgart.de}
\thanks{This work is supported by the German Federal Ministry of Research, Technology and Space (BMFTR) within the project Open6GHub (grant no. 16KISK019).} 
}

\maketitle

\begin{abstract}
Motivated by the need for channel codes with low-complexity soft-decision decoding algorithms, we consider the recursive Plotkin concatenation of optimal low-rate and high-rate codes based on simplex codes and their duals. 
These component codes come with low-complexity \ac{ML} decoding which, in turn, enables efficient \ac{SC}-based decoding.
As a result, the proposed \ac{ORCAS} codes achieve a performance that is at least as good as that of polar codes.
For practical parameters, the proposed construction significantly outperforms polar codes in terms of block error rate by up to 0.5\,dB while maintaining similar decoding complexity. 
Furthermore, the codes offer greater flexibility in codeword length than conventional polar codes.
\end{abstract}

\acresetall

\section{Introduction}\label{sec:intro}
Polar codes \cite{arikan2009} are based on the recursive Plotkin construction \cite{plotkin1960} and are proven to achieve the capacity of binary memoryless channels.
The availability of a low-complexity soft-decision decoder in the form of the \ac{SC} decoding algorithm \cite{schnabl1995rmgcc} makes them attractive channel coding candidates for low-power devices \cite{tong2023fastpolar}.
However, the finite-length \ac{BLER} performance of polar codes under \ac{SC} decoding remains limited due to the suboptimality of both the code construction and the decoding algorithm. 
These shortcomings can be mitigated through the use of an outer code and list decoding \cite{talvardy2015list}, albeit at the cost of significantly increased decoding complexity.
An alternative approach to enhancing polar codes is to employ more capable component codes within the Plotkin concatenation.
In particular, \ac{BCH}-based Plotkin concatenated codes have recently gained attention \cite{goldin2019concatenated, bailon2022bchplotkin, cheng2024bchuuv}.
By viewing the Plotkin concatenation as generalized concatenated codes, the \ac{SC} algorithm can still be applied \cite{trifonov2011gcc, schnabl1995rmgcc}. 
Nevertheless, the component \ac{BCH} codes typically require complex soft-decision decoding such as \ac{OSD} and are not very flexible in code rate and length.
Specifically, the overall code length is constrained to powers of two times the component-code length.
While there exist length-matching procedures such as shortening or puncturing for polar-like codes, they reduce the \ac{BLER} performance.

\hspace{-0.35pt}In this paper, we propose \underline{o}ptimally \underline{r}ecursively con\-\underline{ca}t\-e\-nated \underline{s}implex (ORCAS)\acused{ORCAS} codes, which employ near-optimal low-rate and optimal high-rate component codes in the recursive Plotkin concatenation. 
Medium-rate nodes are recursively split until optimal component codes are obtained.
For these component codes, which are based on simplex codes and their duals, low-complexity (near-) \ac{ML} decoders exist, resulting in a decoding complexity similar to that of \ac{SC} decoding of polar codes.
\Ac{ORCAS} codes are inherently flexible in their length and rate, making them particularly attractive for block lengths that are not powers of two.
As \ac{ORCAS} codes generalize polar codes, their \ac{BLER} performance is at least as good as that of polar codes with \ac{SC} decoding, and typically up to $0.5\,\mathrm{dB}$ better.

\section{Preliminaries}\label{sec:prelim}

\subsection{Notations}
For a non-negative integer $i$, let $\boldsymbol{b}_{m,i} = (i_0, i_1, \dots, i_{m-1})$ denote its $m$-bit \ac{LSB}-first binary vector expansion with bit-length $l_\mathrm{b}(i) = \lceil \log_2(i+1) \rceil$. 
The \text{$m$-bit} binary expansion matrix of integers $1$ to $n$ is 
\begin{equation*}
    \boldsymbol{B}_{m,n} = 
    \begin{bmatrix}
        \boldsymbol{b}_{m,1}\tp & \boldsymbol{b}_{m,2}\tp & \dots & \boldsymbol{b}_{m,n}\tp
    \end{bmatrix} \in \mathbb{F}_2^{m \times n}.
\end{equation*}
For a vector $\boldsymbol{v} \in \mathbb{F}_2^n$, $\w(\boldsymbol{v})= |\operatorname{supp}(\boldsymbol{v})|$ gives its Hamming weight.
We use the shorthand notation $[i,j] = \{i,i+1,\dots,j\}$, 
where each endpoint is included if written with a bracket and excluded if written with a parenthesis.
The $i$-th Mersenne number is denoted as $M_i = 2^i-1$.

\subsection{Binary Linear Block Codes}
A binary linear code $\mathcal{C} \subseteq \mathbb{F}_2^n$ with parameters $(n,k,d)$ has block length $n$, dimension $k$, minimum distance $d$, and rate $R = k/n$.
It can be described by a full-row-rank generator matrix $\boldsymbol{G} \in \mathbb{F}_2^{k \times n}$ or parity-check matrix $\boldsymbol{H} \in \mathbb{F}_2^{(n-k) \times n}$ as 
\begin{equation*}
    \mathcal{C} = \mleft\{\boldsymbol{u} \boldsymbol{G} \,\middle\vert\, \boldsymbol{u} \in \mathbb{F}_2^k\mright\} = \mleft\{\boldsymbol{c} \in \mathbb{F}_2^n \,\middle\vert\, \boldsymbol{H} \boldsymbol{c}\tp = \boldsymbol{0}\mright\},    
\end{equation*}
with $\boldsymbol{G} \boldsymbol{H}\tp = \boldsymbol{0}$.
The minimum distance of $\mathcal{C}$ is given by
\begin{equation*}
    d(\mathcal{C}) = \min_{\boldsymbol{c} \in \mathcal{C} \setminus \{\boldsymbol{0}\}} \w(\boldsymbol{c}).
\end{equation*}
The dual code $\smash{\mathcal{C}}^\perp$ is defined as the set of all vectors orthogonal to $\mathcal{C}$ and is generated by $\boldsymbol{G}^\perp = \boldsymbol{H}$.

\subsection{\rule{0pt}{2.9ex}Plotkin Construction, Successive Cancellation Decoding, and Polar Codes}
The recursive Plotkin construction~\cite{plotkin1960} provides a systematic method to build longer and stronger codes from shorter ones.
Given two component codes $\mathcal{C}_1$ $(n,k_1,d_1)$ and $\mathcal{C}_2$ $(n,k_2,d_2)$, 
it combines them into an $(2n,k_1+k_2,\min\{d_1,2d_2\})$ concatenated code 
$\mathcal{C} = \bigl\{ (\boldsymbol{u}\oplus\boldsymbol{v} \mid \boldsymbol{v}) \;\big\vert\; \boldsymbol{u}\in \mathcal{C}_1, \boldsymbol{v}\in \mathcal{C}_2\bigr\}$.

Codes based on the recursive Plotkin construction can be decoded using the \ac{SC} algorithm \cite{schnabl1995rmgcc}. In \ac{SC} decoding, the channel \ac{LLR} vector is split in two halves $\boldsymbol{\ell} = (\boldsymbol{\ell}_1 \mid \boldsymbol{\ell}_2) $ and the input \acp{LLR} $\boldsymbol{\ell}_1'$ for $\mathcal{C}_1$ are computed as
\begin{equation}\label{eq:scf}
    \ell_{1,i}' = \ell_{1,i} \boxplus \ell_{2,i} \approx \operatorname{sgn}(\ell_{1,i})\operatorname{sgn}(\ell_{2,i}) \min \{ |\ell_{1,i}|, |\ell_{2,i}| \},
\end{equation}
where $a \boxplus b = \ln\bigl( \frac{1+\operatorname{e}^{a+b}}{\operatorname{e}^{a}+\operatorname{e}^{b}}\bigr)$.
Then, $\mathcal{C}_1$ is decoded, resulting in the estimate $\hat{\boldsymbol{u}}$. 
This is used to compute the \acp{LLR} $\boldsymbol{\ell}_2'$ for $\mathcal{C}_2$~as
\begin{equation}\label{eq:scg}
    \ell_{2,i}' = (-1)^{\hat{u}_i} \ell_{1,i} + \ell_{2,i}.
\end{equation}
The decoding of $\mathcal{C}_2$ results in $\hat{\boldsymbol{v}}$, with which
the final codeword estimate $\hat{\boldsymbol{c}}=(\hat{\boldsymbol{u}}\oplus\hat{\boldsymbol{v}} \mid \hat{\boldsymbol{v}})$ is obtained.
A block error occurs if $\hat{\boldsymbol{u}} \neq \boldsymbol{u}$ or $\hat{\boldsymbol{v}} \neq \boldsymbol{v}$.  
Let $p_1$ be the \ac{BLER} of $\mathcal{C}_1$ and $p_2$ the one of $\mathcal{C}_2$ for given $\boldsymbol{u}$.
Then, the \ac{BLER} is upper-bounded by
\begin{equation}\label{eq:scbler}
    p_\mathrm{e} \leq 1-(1-p_1)(1-p_2) = p_1 + p_2 - p_1 p_2.
\end{equation}
The upper bound arises because $p_2$ also includes error events with 
$\hat{\boldsymbol{u}} \neq \boldsymbol{u}$, which are counted twice \cite{scbler}.

Under \ac{SC} decoding, the component-code channels asymptotically polarize to either completely noisy or noiseless.
Polar codes \cite{arikan2009} of length $n=2^m$ exploit this by applying the Plotkin construction $m$ times to length-1 component codes (synthetic bit channels), assigning rate-1 to the indices in the information set $\mathcal{I}$, 
which carry the information bits, and rate-0 to the remaining indices, called frozen bits.
Using the binary Hadamard matrix $\boldsymbol{G}_n = \begin{bsmallmatrix}1 & 0 \\ 1 & 1\end{bsmallmatrix}^{\otimes m}$, a codeword is $\boldsymbol{c} = \boldsymbol{u}\boldsymbol{G}_n$, where $\boldsymbol{u}_\mathcal{I} = \boldsymbol{m} \in \mathbb{F}_2^k$ and zero elsewhere.
A method for selecting $\mathcal{I}$ to minimize the \ac{SC} \ac{BLER} is described below.

\subsection{Density Evolution / Gaussian Approximation}
We consider \ac{BPSK} modulation with $X \in \{\pm 1\}$ and transmission over the \ac{BI-AWGN} channel $Y = X + Z$, where $Z \sim \mathcal{N}(0, \sigma^2)$. 
The channel \ac{LLR} is given by $L = \frac{2Y}{\sigma^2}$ and the \ac{SNR} is $E_\mathrm{s}/N_0 = \frac{1}{2\sigma^2}$.

\Ac{DE} is used to analyze the performance of message-passing decoders, such as the \ac{SC} algorithm \cite{scbler, richardson2008}. 
Without loss of generality, we assume the transmission of the all-zero codeword.
Then, the channel \ac{LLR} is distributed as
$[L\mid X=+1] \sim \mathcal{N}\mleft(\tfrac{2}{\sigma^2},\tfrac{4}{\sigma^2}\mright)$, which is fully described by the noise variance $\sigma^2$ of the channel, and equivalently by the mean $\mu =\operatorname{E}[L\mid X=+1] = \nicefrac{2}{\sigma^2}$.

Although \ac{DE} provides accurate performance predictions, it is computationally intensive.
To reduce the complexity, \ac{DEGA} approximates the \ac{LLR} densities propagated through the decoding algorithm as Gaussian distributions 
\rule{0pt}{2.9ex}%
$\mathcal{N}(\mu,2\mu)$
\cite{dega_org}.
The evolution of the distribution mean $\mu$ through \eqref{eq:scf} is given by
\begin{equation}\label{eq:degacn}
    f\mleft(\mu\mright) = \phi^{-1}\mleft(1 - \mleft(1-\phi(\mu)\mright)^2\mright) \approx S(\mu),
\end{equation}
where 
\begin{equation*}
    \phi(x) = 
    \begin{cases}
        1 - \frac{1}{\sqrt{4\pi x}}
        \displaystyle\int_{-\infty}^{\infty}
        \tanh\mleft(\frac{u}{2}\mright)
        e^{-\frac{(u-x)^2}{4x}} \mathop{}\!\mathrm{d}u& \text{if } x > 0,\\
        1 & \text{if } x = 0,
    \end{cases}
\end{equation*}
and $S(\cdot)$ is an accurate and efficient approximation given in \cite[(28)]{oliveira2022}.
The evolution of the mean $\mu$ through \eqref{eq:scg} is found~as
\begin{equation}\label{eq:degavn}
    g\mleft(\mu\mright) = 2\mu.
\end{equation}
After evolving the \ac{LLR} mean $\mu=\nicefrac{2}{\sigma^2}$ through the decoder, the \ac{BER} of a component-code channel under \ac{SC} decoding can be approximated as ${p_\mathrm{b} \approx Q(\nicefrac{1}{\sigma})}$, where $Q(\cdot)$ denotes the complementary \ac{CDF} of $\mathcal{N}(0,1)$.

For polar codes, the information set $\mathcal{I}$ consists of the indices of the synthetic bit channels with the smallest $p_b$.
The \ac{BLER} under \ac{SC} decoding can be upper-bounded by setting $p_b=0$ for the frozen bits and then recursively applying~\eqref{eq:scbler}.

\subsection{Error Probability of Binary Linear Block Codes}
The \ac{BLER} for transmission over a \ac{BI-AWGN} channel using \ac{ML} decoding of an $(n,k,d)$ binary linear block code $\mathcal{C}$ can be upper-bounded by the union bound \cite{unionBound} as 
\begin{equation}\label{eq:unionbound}
    p_\mathrm{e} \leq \overline{p}_\mathrm{e}(\sigma, \{A_w\}_{w=0}^n) = \sum\limits_{w=d}^n A_w 
    Q\mleft(\frac{\sqrt{w}}{\sigma}\mright),
\end{equation}
where the number of codewords $\boldsymbol{c} \in \mathcal{C}$ of Hamming weight~$w$ is denoted as
\begin{equation*}
    A_w = \mleft\vert \mleft\{\boldsymbol{c} \in \mathcal{C} \,\middle\vert\, \w(\boldsymbol{c}) = w \mright\}\mright\vert,\quad 0 \leq w \leq n,
\end{equation*} 
and the ordered set $\{A_w\}_{w=0}^n$ is the weight distribution of $\mathcal{C}$.
At high \ac{SNR}, the contribution of high-weight codewords in~(\ref{eq:unionbound}) becomes negligible, so the bound is well approximated by the lowest-weight terms. 
The minimum distance $d$ and the number of minimum-weight codewords $A_d$ are therefore critical for the \ac{ML} performance.

\section{Optimality of Low-Rate and High-Rate Codes}
We distinguish two notions of optimality to assess code quality.
A code is \emph{distance-optimal} if it achieves the maximal minimum distance, and \emph{\acf{ABLER}-optimal} if it also minimizes the number of minimum-weight codewords.  
\ac{ABLER}-optimality yields the minimum possible \ac{BLER} as $E_\mathrm{s}/N_0 \to \infty$, thereby minimizing the union bound~\eqref{eq:unionbound}.

\begin{definition}[Low-rate code]
An $(n,k,d)$ binary linear block code is said to be low-rate if $k \leq \lb(n)$. It is further called optimal if it is distance-optimal.  
\end{definition}

\begin{definition}[High-rate code]
An $(n,k,d)$ binary linear block code is said to be high-rate if $k \geq n - \lb(n)$. It is further called optimal if it is \ac{ABLER}-optimal.
\end{definition}

By these definitions, the maximal-rate instance of an optimal low-rate code is the first-order \ac{RM} code with $k=\lb(n)$ and $d=\tfrac{n}{2}$.
Further, the minimal-rate instance of an optimal high-rate code is the extended Hamming code with $k=n-\lb(n)$ and~$d=4$.

We find that the low-rate and high-rate component codes of polar codes are sub-optimal, except for first-order \ac{RM} codes and some of their subcodes, extended Hamming codes, and trivial codes.
Furthermore, for \ac{DE}-based design, \ac{SC} decoding does not achieve \ac{ML} performance for $3 < k < n-3$.

These observations motivate the introduction of two alternative component-code classes, which are optimal for most parameters and can be (near-) \ac{ML} decoded with complexity comparable to \ac{SC}.

\section{Near Optimal Low-Rate Codes:\\Naturally Punctured Repeated Simplex Codes}
\subsection{Construction and Optimality}
The $(rM_k,k,r2^{k-1})$ repeated simplex code $\mathcal{S}_k(r)$ is a dis\-tance-optimal code constructed by repeating a binary simplex code $r$ times.
Its generator matrix is formed by concatenating $r$ copies of $\boldsymbol{B}_{k, M_k}$.
A large class of optimal low-rate codes can be constructed by algebraic puncturing of repeated simplex codes \cite{Solomon1965}.\footnote{In fact, any binary linear block code is equivalent to a punctured repeated simplex code $S_k(r)$ for sufficiently large $r$, as its generator matrix contains all possible binary vectors.}
However, the irregular algebraic puncturing pattern complicates the decoder implementation.

Thus, we propose to simplify the puncturing.
To obtain a low-rate code of length $n$ and dimension $k$, the first
\begin{equation*}
    a(n,k) = (-n) \bmod M_k
\end{equation*}
bits of $\mathcal{S}_k\mleft(\lceil \nicefrac{n}{M_k}\rceil\mright)$ are punctured, yielding a \acf{NPRS} code.
Therefore, we give up optimality for some parameters compared to the codes of \cite{Solomon1965}.
\begin{theorem}
    An \ac{NPRS} code of length $n$ is distance-optimal for dimensions $k \in \kappa_n(\lb(n))$, where
    \begin{equation*}
        \kappa_n(k)
        = \begin{cases}
            \{1\} & \text{if } k = 1,\\
            \kappa_n(k-1) & \text{if } a(n,k) \in \mathcal{A}(n,k),\\
            \kappa_n(k-1) \cup \{k\} &  \text{otherwise,}
        \end{cases}
    \end{equation*}
    and $\mathcal{A}(n,k) = \{0\}\cup\mleft\{2^{i}-i,\dots,2^{i}\mright\}$, $i=\lb(a(n,k)-1)$. 
\end{theorem}
\begin{IEEEproof}
    Let $0 \leq i < k$ denote the smallest integer such that $2^{i} \ge a(n,k)$.  
    When $a(n,k) \in \mleft\{ M_{i},\, 2^{i} \mright\}$, an \ac{NPRS} code is equivalent to the distance-optimal codes constructed in~\cite{Solomon1965}.  
    If $a(n,k) = M_{i}$ and $i>0$, then from the Griesmer bound \cite{griesmer_bound} follows that increasing the minimum distance by one requires increasing the length by at least $i$.
    Thus, an \ac{NPRS} code is distance-optimal when $a(n,k) \in \mathcal{A}(n,k)$.
\end{IEEEproof}

For encoding, an im\-ple\-men\-ta\-tion-friendly sorted generator matrix is obtained by repeating each column $\boldsymbol{b}_{k,i}$ of $\boldsymbol{B}_{k,M_k}$
\begin{equation}\label{eq:repetitions}
    \rho_{n,k}(i) 
    = \mleft\lfloor\frac{n}{M_k}\mright\rfloor + 
    \begin{cases}
        1 & \text{if } i > M_k - (n \bmod M_k),\\
        0 &  \text{otherwise,}
    \end{cases}    
\end{equation}
times, where $1 \leq i \leq M_k$.
As discussed in Section~\ref{subs:lowratedec}, \ac{NPRS} codes can be efficiently \ac{ML} decoded using the \ac{FHT}.

\subsection{Weight Distribution}
To evaluate the performance of \ac{NPRS} codes, we determine their weight distribution via the associated \emph{anticode}.
Puncturing a code $\mathcal{C}$ with generator matrix $\boldsymbol{G}$ at positions $\mathcal{P}$ yields a code over $\mathcal{P}^\mathsf{c} = [0,n) \setminus \mathcal{P}$ with generator matrix $\boldsymbol{G}_{:,\mathcal{P}^\mathsf{c}}$.
The removed columns $\boldsymbol{G}_{:,\mathcal{P}}$ form an anticode whose rows may be dependent, its minimum distance may be zero, and each codeword appears $2^{k-r}$ times, where $r = \operatorname{rank}(\boldsymbol{G}_{:,\mathcal{P}})$.
For an \ac{NPRS} code of length $n$ and dimension $k$, we set $\mathcal{P} = [0,a(n,k))$, and the anticode is generated by $\boldsymbol{B}_{k,a(n,k)}$.

The weight distribution of a binary linear block code or anticode with generator matrix $\boldsymbol{G} \in \mathbb{F}_2^{k \times n}$ can be computed by decomposing it into $k+1$ cosets,
\begin{equation*}%
    A_w^{(i)}(\boldsymbol{G}) 
    = \Bigl\lvert\Bigl\{
        \mathcal{J}\subseteq (i,k) \;\Big\vert\; \w\bigl(\boldsymbol{g}_{i} \oplus \bigoplus_{j \in \mathcal{J}} \boldsymbol{g}_j\bigr) = w
    \Bigr\}\Bigr\rvert,
\end{equation*}
for $0 \leq i < k$, where each coset contributes $2^{k-i-1}$ (possibly repeated) codewords.
For $i = k$, we have $A_w^{(k)} = 1$ if $w = 0$ and $0$ otherwise.
The weights appearing in coset $i$ are
\begin{equation*}
    \mathcal{W}_i(\boldsymbol{G}) 
    = \mleft\{w\in[0,n]\;\middle\vert\; A_w^{(i)}(\boldsymbol{G}) > 0\mright\}.
\end{equation*}
The distinct weights in coset $i$ of an anticode generated by $\boldsymbol{B}_{k,a(n,k)}$ are obtained via the following lemma.

\begin{lemma}\label{lem:coset weights}
The coset $i$ of the anticode generated by $\boldsymbol{B}_{k,a}$ has codeword weights $\mathcal{W}_i(\boldsymbol{B}_{k,a}) = \mathcal{V}_i(a+1)$, where
\begin{equation*}
    \mathcal{V}_i(b) = 
    \begin{cases}
        \{w_i(b),b-w_i(b)\} & \text{if } 2^{i+1} \leq b,\\
        \{w_i(b)\} & \text{otherwise},
    \end{cases}
\end{equation*}
and 
$w_i(b) = \mleft\lfloor\nicefrac{b}{2^{i+1}}\mright\rfloor \cdot 2^{i} + \max\mleft\{0,\,\mleft(b \bmod 2^{i+1}\mright)- 2^{i}\mright\}$.
\end{lemma}
\begin{IEEEproof}
    See \appendixname~\ref{app:lem:coset weights}.
\end{IEEEproof}

Using Lemma~\ref{lem:coset weights}, we obtain the weight distribution of an \ac{NPRS} anticode generated by $\boldsymbol{B}_{k,a(n,k)}$.

\begin{theorem}\label{thm:anticode weight dist}
    The anticode generated by $\boldsymbol{B}_{k,a}$ has the weight distribution 
    $A_w\mleft(\boldsymbol{B}_{k,a}\mright) = \sum_{i=0}^{k} A_w^{(i)}(\boldsymbol{B}_{k,a})$,
    where
    \begin{equation*}
        A_w^{(i)}\mleft(\boldsymbol{B}_{k,a}\mright) = 
        \begin{cases}
            2^{k-i-\mleft\lvert\mathcal{W}_i(\boldsymbol{B}_{k,a})\mright\rvert} & \text{if } w \in \mathcal{W}_i(\boldsymbol{B}_{k,a}),\\
            0 & \text{otherwise},
        \end{cases}
    \end{equation*}
    for $0 \leq i < k$ and $A_w^{(k)}=1$ if $w=0$ and $0$ otherwise.
\end{theorem}
\begin{IEEEproof}
    See \appendixname~\ref{app:thm:anticode weight dist}.
\end{IEEEproof}

For any message $\boldsymbol{m} \in \mathbb{F}_2^k$, the weight of the corresponding codeword of a code punctured at positions $\mathcal{P}$ can be found as
\begin{equation}\label{eq:weight by anti}
    \w(\boldsymbol{m}\boldsymbol{G}_{:,\mathcal{P}^\mathsf{c}}) = \w(\boldsymbol{m}\boldsymbol{G}) - \w(\boldsymbol{m}\boldsymbol{G}_{:,\mathcal{P}}).
\end{equation}
With this and Theorem~\ref{thm:anticode weight dist}, the weight distribution of an \ac{NPRS} code can be derived from the corresponding anticode.

\begin{corollary}\label{cor:NPRS weight dist}
    An \ac{NPRS} code of length $n$ and dimension $k$ has the weight distribution
    \begin{equation*}
        A_w^\mathrm{NPRS}(n,k)
        = \begin{cases}
            1 & \text{if } w = 0,\\
            \sum_{i=0}^{k-1} A_{d-w}^{(i)}\mleft(\boldsymbol{B}_{k,a(n,k)}\mright) & \text{otherwise}.
        \end{cases}
    \end{equation*}
    where $d = \lceil \nicefrac{n}{M_k}\rceil \cdot 2^{k-1}$.
\end{corollary}
\begin{IEEEproof}
    See \appendixname~\ref{app:cor:NPRS weight dist}.
\end{IEEEproof}

\section{Optimal High-Rate Codes: NPRS Dual Codes}
\subsection{Construction and Optimality}
By constructing \acf{NPRSD} codes, we obtain optimal high-rate codes.
\begin{theorem}
    An \ac{NPRSD} code of length $n$ is \ac{ABLER}-optimal for dimensions $k \geq n-\lb(n)$.
\end{theorem}
\begin{IEEEproof}
    Consider the following two cases: 
    
    \emph{1) \hspace{-0.06pt}$k = n-\lb(n)$:\hspace{-0.03pt}} 
    The \ac{NPRSD} code is a shortened Hamming code with $d\geq3$ and a parity-check matrix that corresponds to the binary expansions of $M_{n-k}-n$ to $M_{n-k}$. 
    Thus, from \cite[Corollary 1]{HammingOpt} follows that $A_3$ is minimal.
    For $n = 2^{k-1}$, we obtain an extended Hamming code with $d=4$.
    
    \emph{2) $n-\lb(n) < k < n$:}
    Since we have $n > M_{n-k}$, the pa\-ri\-ty-check matrix of the \ac{NPRSD} code contains duplicated non-zero columns and thus $d = 2$. 
    Natural puncturing minimizes the number of duplicated columns and thus minimizes $A_2$.
\end{IEEEproof}

For encoding, we can construct a decoder implementation-friendly generator matrix 
by forming a systematic parity-check matrix from an \ac{NPRS} generator matrix obtained via~\eqref{eq:repetitions}.
As we will see in Section~\ref{subs:highratedec}, \ac{NPRSD} codes can be efficiently near-\ac{ML} decoded using Chase-II syndrome decoding.  

\subsection{Weight Distribution}
To analyze the performance of \ac{NPRSD} codes, we require their weight distribution.  
Using the MacWilliams identity~\cite{macwilliams1977}, 
we can compute the weight distribution of a dual code $\mathcal{C}^\perp$ from that of an $(n,k,d)$ code $\mathcal{C}$ as 
\begin{equation*}%
\begin{aligned}
    A_w^\perp\mleft(\{A_w\}_{w=0}^n\mright)%
    = 2^{-k}  
    \sum_{i=0}^n A_i  \!\!\!
    \sum_{j=\max\{0,i+w-n\}}^{\min\{i,w\}} \!\!\!\!\!\!\!\!\!\!\!(-1)^j\binom{i}{j}\binom{n-i}{w-j}.
\end{aligned}
\end{equation*}
Accordingly, the weight distribution of an \ac{NPRSD} code is obtained from the corresponding \ac{NPRS} code via
\begin{equation*}
    A_w^\mathrm{NPRSD}(n,k)
    = A_w^\perp\mleft(\{A_w^\mathrm{NPRS}(n-k,k)\}_{w=0}^n\mright).
\end{equation*}

\section{Design}
\begin{figure}[t]
    \vspace{0.03cm}
	\centering
	\resizebox{0.95\columnwidth}{!}
    {\begin{tikzpicture}[y=0.75cm,>=stealth,
    tick/.pic = {
    		\draw[line width=0.5pt] (-0.4mm,-0.8mm) -- (0.4mm,0.8mm);
    }
  ]

\tikzstyle{normalnode} = [dspnodefull,minimum size=1mm]
\tikzstyle{VN} = [circle, draw, minimum size = 3mm]
\tikzstyle{dec} = [draw, minimum width = 3mm]
\tikzstyle{CN} = [draw, minimum size = 3mm]
\tikzstyle{dspline} = [line width = 0.25mm]

\tikzstyle{decoder} = [
    fill=gray!10,
    line width=0.5pt,
    minimum width=1.8cm,
]
\tikzstyle{decoder1} = [
    decoder,
    minimum height=0.6cm, 
]
\tikzstyle{decoder2} = [
    decoder,
    minimum height=1.35cm, 
]

\node[decoder2,draw=pink!40, fill=pink!20] (u1) at (0.5,6.5) {$(24, 2,16)$};
\node[decoder1,draw=mittelblau!40, fill=mittelblau!20] (u2) at (-1,5) {$(12, 3, 6)$};
\node[decoder1,draw=apfelgruen!40, fill=apfelgruen!20] (u3) at (-1,4) {$(12, 8, 3)$};
\node[decoder1,draw=mittelblau!40 ,fill=mittelblau!20] (u4) at (-1,3) {$(12, 4, 6)$};
\node[decoder1,draw=apfelgruen!40, fill=apfelgruen!20] (u5) at (-1,2) {$(12, 9, 2)$};
\node[decoder2,draw=mittelgrau!40, fill=mittelgrau!20] (u7) at (0.5,.5) {$(24,22, 2)$};

\node[dspadder] (node2) at (1.00, 5) {};
\node[normalnode] (node3) at (1.00, 4) {};
\draw[dspline,-] (u2)--(node2);
\draw[dspline] (u3)--(node3);
\draw[dspline,-] (node3)--(node2);
\node[dspadder] (node4) at (1.00, 3) {};
\node[normalnode] (node5) at (1.00, 2) {};
\draw[dspline,-] (u4)--(node4);
\draw[dspline] (u5)--(node5);
\draw[dspline,-] (node5)--(node4);

\node[dspadder] (node8) at (3.00, 7) {};
\node[normalnode] (node9) at (3.00, 5) {};
\draw[dspline,-] (u1.east |- node8)--(node8);
\draw[dspline] (node2)--(node9);
\draw[dspline,-] (node9)--(node8);
\node[dspadder] (node10) at (3.50, 6) {};
\node[normalnode] (node11) at (3.50, 4) {};
\draw[dspline,-] (u1.east |- node10)--(node10);
\draw[dspline] (node3)--(node11);
\draw[dspline,-] (node11)--(node10);
\node[dspadder] (node12) at (3.00, 3) {};
\node[normalnode] (node13) at (3.00, 1) {};
\draw[dspline,-] (node4)--(node12);
\draw[dspline,-] (u7.east |- node13)--(node13);
\draw[dspline,-] (node13)--(node12);
\node[dspadder] (node14) at (3.50, 2) {};
\node[normalnode] (node15) at (3.50, 0) {};
\draw[dspline,-] (node5)--(node14);
\draw[dspline] (u7.east |- node15)--(node15);
\draw[dspline,-] (node15)--(node14);
\node[dspadder] (node16) at (5.50, 7) {};
\node[normalnode] (node17) at (5.50, 3) {};
\draw[dspline,-] (node8)--(node16);
\draw[dspline] (node12)--(node17);
\draw[dspline,-] (node17)--(node16);
\node[dspadder] (node18) at (6.00, 6) {};
\node[normalnode] (node19) at (6.00, 2) {};
\draw[dspline,-] (node10)--(node18);
\draw[dspline] (node14)--(node19);
\draw[dspline,-] (node19)--(node18);
\node[dspadder] (node20) at (6.50, 5) {};
\node[normalnode] (node21) at (6.50, 1) {};
\draw[dspline,-] (node9)--(node20);
\draw[dspline] (node13)--(node21);
\draw[dspline,-] (node21)--(node20);
\node[dspadder] (node22) at (7.00, 4) {};
\node[normalnode] (node23) at (7.00, 0) {};
\draw[dspline,-] (node11)--(node22);
\draw[dspline] (node15)--(node23);
\draw[dspline,-] (node23)--(node22);
\node[normalnode] (x0) at (9.50, 7) {};
\draw[dspline,-] (node16)--(x0);
\node[normalnode] (x1) at (9.50, 6) {};
\draw[dspline,-] (node18)--(x1);
\node[normalnode] (x2) at (9.50, 5) {};
\draw[dspline,-] (node20)--(x2);
\node[normalnode] (x3) at (9.50, 4) {};
\draw[dspline,-] (node22)--(x3);
\node[normalnode] (x4) at (9.50, 3) {};
\draw[dspline,-] (node17)--(x4);
\node[normalnode] (x5) at (9.50, 2) {};
\draw[dspline,-] (node19)--(x5);
\node[normalnode] (x6) at (9.50, 1) {};
\draw[dspline,-] (node21)--(x6);
\node[normalnode] (x7) at (9.50, 0) {};
\draw[dspline,-] (node23)--(x7);

\node[label={[yshift=3.5pt]below: \footnotesize $12$}] (t0) at (8.875, 7) {};
\pic at (t0) {tick};
\node[label={[yshift=3.5pt]below: \footnotesize $12$}] (t1) at (8.875, 6) {};
\pic at (t1) {tick};
\node[label={[yshift=3.5pt]below: \footnotesize $12$}] (t2) at (8.875, 5) {};
\pic at (t2) {tick};
\node[label={[yshift=3.5pt]below: \footnotesize $12$}] (t3) at (8.875, 4) {};
\pic at (t3) {tick};
\node[label={[yshift=3.5pt]below: \footnotesize $12$}] (t4) at (8.875, 3) {};
\pic at (t4) {tick};
\node[label={[yshift=3.5pt]below: \footnotesize $12$}] (t5) at (8.875, 2) {};
\pic at (t5) {tick};
\node[label={[yshift=3.5pt]below: \footnotesize $12$}] (t6) at (8.875, 1) {};
\pic at (t6) {tick};
\node[label={[yshift=3.5pt]below: \footnotesize $12$}] (t7) at (8.875, 0) {};
\pic at (t7) {tick};

\tikzstyle{parameters} = [    
    fill=white, 
    fill opacity=0.67,
    text opacity=1,
    draw=black, 
    rounded corners=3.5pt, 
    line width=0.5pt,
]

\node[parameters, minimum height=0] () at (2,4.50) {\rotatebox{90}{\scriptsize $(24,11,6)$}};
\node[parameters, minimum height=0] () at (2,2.50) {\rotatebox{90}{\scriptsize $(24,13,4)$}};
\node[parameters, minimum height=2.915cm] () at (4.5,5.50) {\rotatebox{90}{\scriptsize $(48,13,12)$}};
\node[parameters, minimum height=2.915cm] () at (4.5,1.50) {\rotatebox{90}{\scriptsize $(48,35,4)$}};
\node[parameters, minimum height=5.915cm] () at (8,3.50) {\rotatebox{90}{\scriptsize $(96,48,8)$}};

\tikzset{
  legendbox/.style={
    rectangle,
    draw,
    minimum width=2.5cm,
    align=center
  }
}

\coordinate (legendanchor) at ([yshift=-0.5cm]current bounding box.south);

\node[legendbox, draw=pink!40, fill=pink!20]     (L1) at ($(legendanchor)+(-4.2,0)$) {CW};
\node[legendbox, draw=mittelblau!40, fill=mittelblau!20]     (L2) at ($(legendanchor)+(-1.4,0)$) {FHT};
\node[legendbox, draw=apfelgruen!40, fill=apfelgruen!20]   (L3) at ($(legendanchor)+(1.4,0)$) {Chase-II};
\node[legendbox, draw=mittelgrau!40, fill=mittelgrau!20] (L4) at ($(legendanchor)+(4.2,0)$) {CW dual};

\begin{scope}[shift={($(legendanchor)+(-4.1,0)$)}]
\node[CN] (cn) at (-1,-4) {};
\node[VN] (vn1) at (0,-2) {};
\node[VN] (vn2) at (0,-4) {};
\node[VN] (vn3) at (0,-6) {};
\node[normalnode] (n11) at (1,-1.25) {};
\node[normalnode] (n12) at (1,-1.75) {};
\node[]           (x1)  at (1,-2.12) {$\vdots$};
\node[normalnode] (n13) at (1,-2.75) {};

\node[normalnode] (n21) at (1,-3.25) {};
\node[normalnode] (n22) at (1,-3.75) {};
\node[]           (x2)  at (1,-4.12) {$\vdots$};
\node[normalnode] (n23) at (1,-4.75) {};

\node[normalnode] (n31) at (1,-5.25) {};
\node[normalnode] (n32) at (1,-5.75) {};
\node[]           (x3)  at (1,-6.12) {$\vdots$};
\node[normalnode] (n33) at (1,-6.75) {};

\draw (cn) -- (vn1);
\draw (cn) -- (vn2);
\draw (cn) -- (vn3);

\draw (vn1) -- (n11);
\draw (vn1) -- (n12);
\draw (vn1) -- (n13);

\draw (vn2) -- (n21);
\draw (vn2) -- (n22);
\draw (vn2) -- (n23);

\draw (vn3) -- (n31);
\draw (vn3) -- (n32);
\draw (vn3) -- (n33);

\end{scope}

\begin{scope}[shift={($(legendanchor)+(-1.3,0)$)}]
\node[dec, rotate=90] (cn) at (-1,-4) {\footnotesize Simplex FHT};
\node[VN] (vn1) at (0,-1.5) {};
\node[VN] (vn2) at (0,-3.5) {};
\node[VN] (vn3) at (0,-6.5) {};
\node[normalnode] (n11) at (1,-0.75) {};
\node[normalnode] (n12) at (1,-1.25) {};
\node[]           (x1)  at (1,-1.62) {$\vdots$};
\node[normalnode] (n13) at (1,-2.25) {};

\node[normalnode] (n21) at (1,-2.75) {};
\node[normalnode] (n22) at (1,-3.25) {};
\node[]           (x2)  at (1,-3.62) {$\vdots$};
\node[normalnode] (n23) at (1,-4.25) {};

\node[normalnode] (n31) at (1,-5.75) {};
\node[normalnode] (n32) at (1,-6.25) {};
\node[]           (x3)  at (1,-6.62) {$\vdots$};
\node[normalnode] (n33) at (1,-7.25) {};

\node[]           (x4)  at (0.5,-5+0.13) {$\vdots$};

\draw (cn) -- (vn1);
\draw (cn) -- (vn2);
\draw (cn) -- (vn3);

\draw (vn1) -- (n11);
\draw (vn1) -- (n12);
\draw (vn1) -- (n13);

\draw (vn2) -- (n21);
\draw (vn2) -- (n22);
\draw (vn2) -- (n23);

\draw (vn3) -- (n31);
\draw (vn3) -- (n32);
\draw (vn3) -- (n33);

\end{scope}

\begin{scope}[shift={($(legendanchor)+(1.5,0)$)}]
\node[dec, rotate=90] (cn) at (-1,-4) {\footnotesize Hamming Chase-II};
\node[CN] (vn1) at (0,-1.5) {};
\node[CN] (vn2) at (0,-3.5) {};
\node[CN] (vn3) at (0,-6.5) {};
\node[normalnode] (n11) at (1,-0.75) {};
\node[normalnode] (n12) at (1,-1.25) {};
\node[]           (x1)  at (1,-1.62) {$\vdots$};
\node[normalnode] (n13) at (1,-2.25) {};

\node[normalnode] (n21) at (1,-2.75) {};
\node[normalnode] (n22) at (1,-3.25) {};
\node[]           (x2)  at (1,-3.62) {$\vdots$};
\node[normalnode] (n23) at (1,-4.25) {};

\node[normalnode] (n31) at (1,-5.75) {};
\node[normalnode] (n32) at (1,-6.25) {};
\node[]           (x3)  at (1,-6.62) {$\vdots$};
\node[normalnode] (n33) at (1,-7.25) {};

\node[]           (x4)  at (0.5,-5+0.13) {$\vdots$};

\draw (cn) -- (vn1);
\draw (cn) -- (vn2);
\draw (cn) -- (vn3);

\draw (vn1) -- (n11);
\draw (vn1) -- (n12);
\draw (vn1) -- (n13);

\draw (vn2) -- (n21);
\draw (vn2) -- (n22);
\draw (vn2) -- (n23);

\draw (vn3) -- (n31);
\draw (vn3) -- (n32);
\draw (vn3) -- (n33);

\end{scope}

\begin{scope}[shift={($(legendanchor)+(4.3,0)$)}]
\node[VN] (cn) at (-1,-4) {};
\node[CN] (vn1) at (0,-2) {};
\node[CN] (vn2) at (0,-4) {};
\node[CN] (vn3) at (0,-6) {};
\node[normalnode] (n11) at (1,-1.25) {};
\node[normalnode] (n12) at (1,-1.75) {};
\node[]           (x1)  at (1,-2.12) {$\vdots$};
\node[normalnode] (n13) at (1,-2.75) {};

\node[normalnode] (n21) at (1,-3.25) {};
\node[normalnode] (n22) at (1,-3.75) {};
\node[]           (x2)  at (1,-4.12) {$\vdots$};
\node[normalnode] (n23) at (1,-4.75) {};

\node[normalnode] (n31) at (1,-5.25) {};
\node[normalnode] (n32) at (1,-5.75) {};
\node[]           (x3)  at (1,-6.12) {$\vdots$};
\node[normalnode] (n33) at (1,-6.75) {};

\draw (cn) -- (vn1);
\draw (cn) -- (vn2);
\draw (cn) -- (vn3);

\draw (vn1) -- (n11);
\draw (vn1) -- (n12);
\draw (vn1) -- (n13);

\draw (vn2) -- (n21);
\draw (vn2) -- (n22);
\draw (vn2) -- (n23);

\draw (vn3) -- (n31);
\draw (vn3) -- (n32);
\draw (vn3) -- (n33);

\end{scope}

\end{tikzpicture}}
	\caption{\footnotesize Top: Factor graph of a $(96,48,8)$ \ac{ORCAS} code with the $(n,k,d)$ parameters of the component codes. Bottom: Factor graphs of the component decoders. Squares and circles denote check and variable nodes, respectively.}
    \label{fig:examplefactorgraph}
\end{figure}
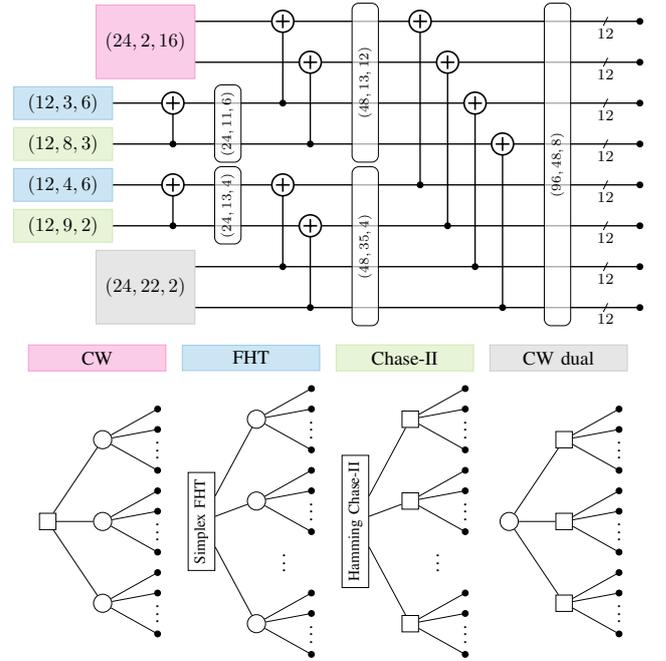

The proposed \ac{ORCAS} codes are constructed via the Plotkin concatenation of \ac{NPRS} and \ac{NPRSD} codes.
Fig.~\ref{fig:examplefactorgraph} illustrates the basic structure of a $(96,48,8)$ \ac{ORCAS} code, consisting of \ac{NPRS} codes $(24,2,16)$, $(12,3,6)$, $(12,4,6)$ and \ac{NPRSD} codes $(12,8,3)$, $(12,9,2)$, $(24,22,2)$.

The code design involves selecting which stages are split with the Plotkin concatenation and assigning specialized component codes to the leaf nodes. %
This is done with Algorithm~\ref{alg:construction}, which is an extension of \ac{DE}.
We describe it using \ac{DEGA} for clarity, though full-precision \ac{DE} with histogram densities of sufficient resolution works similarly.

Starting from the design \ac{SNR} $\nicefrac{1}{2\sigma^2}$, the block length $n$ is recursively halved and the density means $\mu$ are updated via~\eqref{eq:degacn} and~\eqref{eq:degavn}.
For $n\in\{2,3,5,7,9\}$, suitable low-rate \ac{NPRS} and high-rate \ac{NPRSD} codes exist for all dimensions, terminating the recursion.
For each $k=1,\dots,n$, the \ac{BLER} is estimated via the union bound (clipped at the uncoded \ac{BLER}) to form a \ac{BLER} vector $\boldsymbol{p}$, with sorting permutation $\boldsymbol{\pi}$.

Unlike standard \ac{DE}, we apply an upward recursion that combines the component-code \ac{BLER} vectors $\boldsymbol{u}$ and $\boldsymbol{v}$ via~\eqref{eq:scbler}.
Using the sorting permutations, we iteratively identify whether increasing component-code dimension $k_1$ or $k_2$ yields a lower overall \ac{BLER}.
Thus, for every total dimension $k=k_1+k_2$, the recursion returns the minimal \ac{BLER} among all admissible $(k_1,k_2)$.
As a result, this procedure reduces the combination in lines \textbf{\small\ref{algoline:start}} to \textbf{\small\ref{algoline:end}} from quadratic to linear time.
The selections are \emph{optimal} when~\eqref{eq:scbler} is tight, which holds at medium-to-low \acp{BLER}.
Finally, for each $(n,k)$ admitting a suitable \ac{NPRS} or \ac{NPRSD} code, we retain the lower value between its union bound and the \ac{BLER} of the concatenated code.

Since both \eqref{eq:scbler} and the union bound \eqref{eq:unionbound} are upper bounds, the algorithm with exact \ac{DE} yields, for each $k$, an upper bound on the \ac{BLER} of the (optimal) component-code combination at the design \ac{SNR}.
The resulting \ac{ORCAS} code is specified by a rate-profile vector $\boldsymbol{r}$, where the positions of the $k$ smallest entries in $\boldsymbol{p}$ are set to~1 and the remaining positions to~0.
In line with the Plotkin construction, $\boldsymbol{r}$ is interpreted blockwise, where the block size gives the component-code length and the number of 1s its dimension.

\mbox{}\vspace{-2.25em} %
\begin{algorithm}[ht]
    \caption{\footnotesize Construction of \ac{ORCAS} codes using \ac{DEGA}.} 
    \label{alg:construction}
    \small
    \SetAlgoLined\LinesNumbered
    \SetKwInOut{Input}{Input}\SetKwInOut{Output}{Output}
    \Input{\ac{SNR} $E_\mathrm{s}/N_0$, code length $n$, code dimension $k$.}
    \Output{Rate-profile $\boldsymbol{r}$.}
    $-, \boldsymbol{\pi} \gets \operatorname{evolve}(4 \cdot \nicefrac{E_\mathrm{s}}{N_0},n)$\;
    $\boldsymbol{r} \gets \boldsymbol{0}$; 
    $r_{\pi_{i}} \gets 1 \quad \forall \, i = 0, \dots, k-1$\; 
    \algrule[.5pt]
    \Fn{$\operatorname{evolve}(\mu,\,n)$}{
    $i,\, j,\,\overline{u},\,\overline{v} \gets 0$; 
    $k \gets 1$; 
    $\boldsymbol{p},\, \boldsymbol{\pi} \gets \boldsymbol{0}$\;%
    \If{$n \bmod 2 = 1$ \rm{\textbf{or}} $n = 2$}{
        $p_{n-k},\,\pi_{n-k} \gets \operatorname{bler}(\mu,n,k),\,k-1 \quad \forall \, k = 1,\dots,n$\;
        \Return $\boldsymbol{p},\, \boldsymbol{\pi}$\;
    }
    $(\boldsymbol{u},\,\boldsymbol{\sigma}), (\boldsymbol{v},\,\boldsymbol{\tau}) \gets \operatorname{evolve}(f(\mu),\nicefrac{n}{2}),\, \operatorname{evolve}(g(\mu),\nicefrac{n}{2})$\;

	\While{$i < \nicefrac{n}{2}$ \rm{\textbf{or}} $j < \nicefrac{n}{2}$}{ \label{algoline:start}
        $\overline{p}_1 \gets \overline{v}+u_{\sigma_i} - \overline{v}u_{\sigma_i}$ 
        \rm{\textbf{if}} $i < \nicefrac{n}{2}$ \rm{\textbf{else}} 
        $\overline{p}_1 \gets 1$\; 
        
        $\overline{p}_2 \gets \overline{u}+v_{\tau_j} - \overline{u}v_{\tau_j}$ 
        \rm{\textbf{if}} $j < \nicefrac{n}{2}$ \rm{\textbf{else}} 
        $\overline{p}_2 \gets 1$\; 
        \eIf{$\overline{p}_1 < \overline{p}_2$}{
	        $\overline{u} \gets u_{\sigma_i}$;
	        $\pi_{k-1} \gets \sigma_i$;
	        $i \gets i + 1$\;
	    }{
	        $\overline{v} \gets v_{\tau_j}$;
	        $\pi_{k-1} \gets \tau_j + \nicefrac{n}{2}$; 
	        $j \gets j + 1$\;	    
	    }
	    $p_{\pi_{k-1}} \gets \overline{u} + \overline{v} - \overline{u}\,\overline{v}$; 
	    $k \gets k + 1$\;
	} \label{algoline:end}
    $\mathcal{K} \gets \{2,\dots,\lb(n)\}\cup\{n-\lb(n),\dots,n-2\}$\;
	$p_{\pi_{k-1}} \gets \min\mleft\{p_{\pi_{k-1}}, \operatorname{bler}(\mu,n,k)\mright\}\quad \forall\,k\in\mathcal{K}$\;
	\Return $\boldsymbol{p},\, \boldsymbol{\pi}$\;
	}
	\algrule[.5pt]
    \Fn{$\operatorname{bler}(\mu,\,n,\,k)$}{
        $\sigma \gets \sqrt{\nicefrac{2}{\mu}}$\;
        $\{A_w\}_{w=0}^n \gets \begin{cases}
            \{A_w^\mathrm{NPRS}(n,k)\}_{w=0}^n & \text{if } k \leq \lb(n),\\
            \{A_w^\mathrm{NPRSD}(n,k)\}_{w=0}^n & \text{otherwise};
        \end{cases}$\\
        \Return $\min\bigl\{\overline{p}_\mathrm{e}(\sigma, \{A_w\}_{w=0}^n),1 - (1 - Q(\nicefrac{1}{\sigma}))^n\bigr\}$\;
    }
\end{algorithm}

\section{Decoding}
\subsection{Plotkin Concatenation}
The proposed \ac{ORCAS} codes are recursively decoded using the \ac{SC} algorithm as given in \eqref{eq:scf} and \eqref{eq:scg}.
At the leaf nodes defined by the rate-profile $\boldsymbol{r}$, dedicated component-code decoders are invoked. 
Thus, decoding resembles \ac{SSC} decoding of polar codes \cite{simplifiedSC}, but with different leaf nodes. 
Their factor graphs, shown in the lower part of Fig.~\ref{fig:examplefactorgraph}, are trees and can be optimally decoded by message passing~\cite{richardson2008}. 
The following subsections summarize their decoding.

\subsection{Low-Rate Component Codes}\label{subs:lowratedec}
The low-rate \ac{NPRS} codes can be described as the concatenation of an outer (punctured) simplex code and inner repetition codes.
They are \ac{ML}-decoded by summing the \acp{LLR} of the repeated simplex bits and then applying the \ac{FHT}-based simplex decoder~\cite{GreenFHT}.
\ac{NPRS} codes with $k=2$ reduce to \ac{CW} codes, which are \ac{SPC} codes with repeated bits~\cite{cordaro1967}.
Accordingly, the simplex decoding simplifies to \ac{SPC} decoding.
For $k=1$, we have a repetition code, identically to \ac{SSC} decoding of polar codes.

\subsection{High-Rate Component Codes}\label{subs:highratedec}
The high-rate \ac{NPRSD} codes are decoded analogously. 
In contrast to \ac{NPRS} codes, they can be described as the concatenation of an outer (shortened) Hamming code and inner \ac{SPC} codes.
First, the inner \ac{SPC} codes are soft-combined using the min-ap\-prox\-i\-ma\-tion of the ``$\boxplus$''-operation \eqref{eq:scf}. 
The resulting noisy Hamming codeword is decoded by Chase-II decoding \cite{chase1972}, 
i.e., flipping all subsets of the $p=\lb(n)$ least reliable bits and applying the syndrome decoder. 
Finally, the least reliable bits of the \acp{SPC} inconsistent with the Hamming codeword are flipped.
For $k=n-2$, \ac{NPRSD} codes reduce to \ac{CW} dual codes and the Hamming decoding simplifies to repetition decoding.
For $k=n-1$, we obtain an \ac{SPC} code.

\section{Results}\label{sec:results}
\subsection{Error-Correction Performance}
\begin{figure*}[htp]
	\centering
	\resizebox{\linewidth}{!}{\input{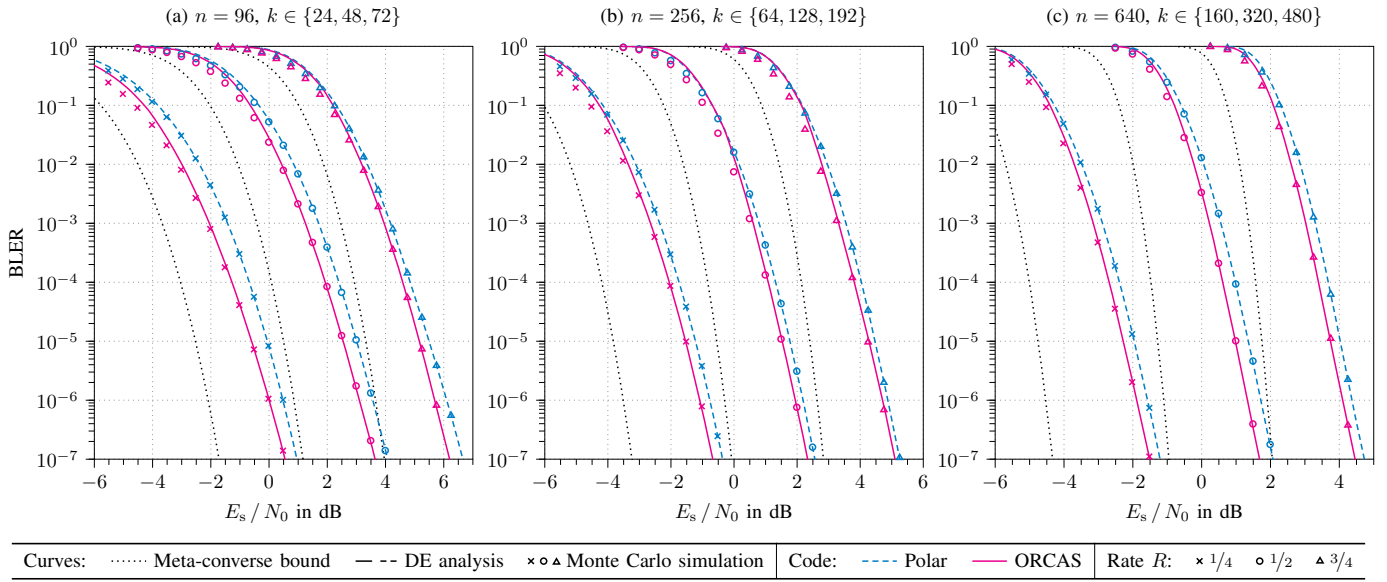}}
	\caption{\footnotesize 
	\Ac{BLER} performance comparison of \ac{ORCAS} codes and polar codes under \ac{SC} decoding. All codes are optimized for a \ac{BLER} of $10^{-6}$ by \ac{DE}.
    }
	\label{fig:blerresults}
\end{figure*}

We evaluate the proposed \ac{ORCAS} codes using Monte Carlo simulation and \ac{DE}, as described in Algorithm~\ref{alg:bler}. 
Like for polar codes, the \ac{SC} decoding performance is upper-bounded by first calculating the error-rate of each component code via \ac{DE} and then estimating the overall \ac{BLER} by recursively applying~\eqref{eq:scbler}.

\begin{algorithm}[ht]
    \caption{\footnotesize \ac{ORCAS} code \ac{BLER} upper bound via \ac{DEGA}.}
    \label{alg:bler}
    \small
    \SetAlgoLined\LinesNumbered
    \SetKwInOut{Input}{Input}\SetKwInOut{Output}{Output}
    \Input{\ac{SNR} $E_\mathrm{b}/N_0$, rate-profile $\boldsymbol{r}$.}
    \Output{\Ac{BLER} upper bound $\overline{p}_\mathrm{e}$.}
    $n \gets \operatorname{length}(\boldsymbol{r})$;
    $k \gets \sum_{i=0}^{n-1} r_i$\;
    \vspace{0.05cm}$\overline{p}_\mathrm{e} \gets \operatorname{evaluate}(4 \cdot \nicefrac{k}{n}\cdot \nicefrac{E_\mathrm{b}}{N_0},\boldsymbol{r})$\;
    \algrule[.5pt]
    \Fn{$\operatorname{evaluate}(\mu,\,\boldsymbol{r})$}{
        $n \gets \operatorname{length}(\boldsymbol{r})$;
        $k \gets \sum_{i=0}^{n-1} r_i$\;
        \lIf{$k = 0$}{\Return $0$}
        \If{$k \leq \lb(n)$ \rm{\textbf{or}} $k \geq n-\lb(n)$}{
            $\boldsymbol{p},\boldsymbol{\pi} \gets (\operatorname{Algorithm}\,\ref{alg:construction}).\operatorname{evolve}(\mu,n)$\;%
            \Return $p_{\pi_{k-1}}$\;
        }
        $\overline{u},\,\overline{v} \gets \operatorname{evaluate}(f(\mu),\boldsymbol{r}_{0:\nicefrac{n}{2}}),\,\operatorname{evaluate}(g(\mu),\boldsymbol{r}_{\nicefrac{n}{2}:n})$\;
        \Return $\overline{u} + \overline{v} - \overline{u}\,\overline{v}$\;
	}
\end{algorithm}

We design \ac{ORCAS} codes for lengths $n\in\{96,256,640\}$, rates $R\in\{\nicefrac{1}{4},\nicefrac{1}{2},\nicefrac{3}{4}\}$, and target \ac{BLER}~$10^{-6}$. 
For comparison, polar codes are constructed with identical parameters.
For block lengths $n=96$ and $n=640$, polar codes are obtained by length matching from $n=128$ and $n=1024$, whereas \ac{ORCAS} codes are constructed natively. 
Table~\ref{tab:polarlengthmatching} lists the applied length-matching techniques.
“Natural” indicates puncturing or shortening from the start or end of the codeword, respectively, whereas “bit-reverse” removes symbols in bit-reversed order starting from the respective codeword end \cite{nui2013lengthmatching}.

\begin{table}[ht]
  \centering\footnotesize
  \caption{\footnotesize Length matching of the baseline polar codes for a \ac{BLER} of $10^{-6}$.}
  \label{tab:polarlengthmatching}
  \begin{NiceTabular}{c|c|ccc}
        \toprule
        Length $n$ & Aspect & $R=\nicefrac{1}{4}$ & $R=\nicefrac{1}{2}$ & $R=\nicefrac{3}{4}$ \\
        \midrule
        \multirow{2}{*}{$96$} 
        & Type & Bit-reverse & Natural & Natural \\
        & Method & Puncturing & Shortening & Shortening \\
        \midrule
        \multirow{2}{*}{$640$} 
        & Type & Natural & Bit-reverse & Natural \\
        & Method & Puncturing & Shortening & Shortening \\
        \bottomrule
  \end{NiceTabular}
\end{table}

Fig.~\ref{fig:blerresults} compares the \ac{BLER} performance of the proposed \ac{ORCAS} codes to polar codes.
Moreover, finite-length performance limits are indicated using the saddlepoint approximations of the \ac{PPV} meta-converse bound \cite{SaddlePointApproxMC}.
\ac{ORCAS} codes show a gain in error-correction performance of $0.3$ to $0.5\,\mathrm{dB}$ over polar codes for all parameters, with the largest gains for lengths $n\neq2^m$.
At low \acp{BLER}, \ac{DE} accurately predicts the actual \ac{BLER}, while at higher \acp{BLER}, \ac{DE} slightly overestimates it because the upper bound~\eqref{eq:scbler} becomes loose.

\subsection{Decoding Complexity}
We assess decoding complexity by measuring the throughput of the proposed \ac{SC}-based \ac{ORCAS} decoder and a \ac{SSC} polar decoder using rate-0, repetition, \ac{SPC}, rate-1, as well as Type-I to Type-IV nodes \cite{SSCtype1to5}.
Both are implemented in C with floating-point arithmetic and executed single-threaded on an Intel Core i9-13900K.
Table~\ref{tab:throughput} lists the achieved throughput in codewords per second across all evaluated lengths and rates.

\begin{table}[htb]
    \centering
    \footnotesize
    \caption{\footnotesize Throughput of \ac{ORCAS} and polar decoders in codewords per second.}
    \label{tab:throughput}
    \begin{NiceTabular}{c|c|ccc}
        \toprule
        Length $n$ & Code & $R=\nicefrac{1}{4}$ & $R=\nicefrac{1}{2}$ & $R=\nicefrac{3}{4}$ \\
        \midrule
        \multirow{2}{*}{96} & Polar & 1,727,526 & 1,281,094 & 1,435,785\\ 
        & \ac{ORCAS} & 1,927,945 & 1,543,126 & 1,509,279 \\
        \midrule
        \multirow{2}{*}{$256$} & Polar & \phantom{0,}692,095 &  \phantom{0,}586,062 & \phantom{0,}604,761 \\ %
        & \ac{ORCAS} & \phantom{0,}763,846 & \phantom{0,}695,437 & \phantom{0,}601,917 \\ %
        \midrule
        \multirow{2}{*}{640} & Polar & \phantom{0,}277,490 & \phantom{0,}225,396 & \phantom{0,}187,966 \\ 
        & \ac{ORCAS} & \phantom{0,}299,271 & \phantom{0,}271,726 & \phantom{0,}317,018\\
        \bottomrule
    \end{NiceTabular}
\end{table}

The results demonstrate that \ac{ORCAS} codes have decoding complexity comparable to polar codes.
Note that for lengths $n \neq 2^m$, the full polar \ac{SC} decoder is applied to the received word with punctured and shortened bits.

\section{Conclusion and Outlook}\label{sec:conclusion}
This paper introduced \ac{ORCAS} codes, a novel coding scheme based on the Plotkin concatenation of efficiently decodable high-rate and low-rate codes.
We proposed a \ac{DE}-based design method to identify optimal \ac{NPRS} and \ac{NPRSD} code combinations.
Simulation and \ac{DE} results show that \ac{ORCAS} codes outperform polar codes under low-complexity \ac{SC} decoding, while also offering greater flexibility in block length than polar and \ac{BCH}-based Plotkin constructions.
Future work includes exploring list decoding for \ac{ORCAS} codes.

\bibliographystyle{IEEEtran}
\bibliography{bibliofile}

\appendices

\renewcommand{\thesubsection}{\Alph{subsection}}

\section*{Appendix}
\subsection{Proof of Lemma~\ref{lem:coset weights}}\label{app:lem:coset weights}
We divide the rows of $\boldsymbol{F} = (\boldsymbol{0}\tp \mid \boldsymbol{B}_{k,a})$ into blocks
\begin{equation*}
    \boldsymbol{g}_{i,j}(m) = \boldsymbol{F}_{i,\, m \, 2^j\, : \,\min\{(m+1) \, 2^j,\,b\}},
\end{equation*}
with $b = a+1$. 
For these, it holds that 
$\boldsymbol{g}_{i,i+1}(m) = (\boldsymbol{b}_{2^i,0} \mid \boldsymbol{1})$ 
and 
$\boldsymbol{g}_{i,j}(m) = \mleft(\lfloor\nicefrac{m}{2^{i-j}}\rfloor\bmod 2\mright) \boldsymbol{1}$ for $i \geq j$.
The $m$-th block of a codeword defined by $\mathcal{J} \subseteq (i,k)$ in coset $i$ has weight
\begin{equation*}
    \tilde{w}_i(m,\mathcal{J})
    = \w\Bigl(\boldsymbol{g}_{i,i+1}(m) 
        \oplus \bigoplus_{j\in \mathcal{J}} \boldsymbol{g}_{j,i+1}(m)\Bigr).
\end{equation*}
It follows that
\begin{equation*}
    \tilde{w}_i(m,\mathcal{J}) 
    \in 
    \begin{cases}
        \{2^i\} & \text{if } m < q_i \text{ and } q_i > 0,\\
        \{s_i, r_i - s_i\} & \text{if } m = q_i \text{ and } q_i > 0,\\
        \{s_i\} & \text{if } m = q_i  \text{ and } q_i = 0,\\
    \end{cases}
\end{equation*}    
for all $\mathcal{J}$, where $q_i = \mleft\lfloor\nicefrac{b}{2^{i+1}}\mright\rfloor$, and \rule{0pt}{3ex}${r_i = b \bmod 2^{i+1}}$, and $s_i = \max\mleft\{0,r_i - 2^{i}\mright\}$.
Thus, coset $i$ has the codeword weights
\begin{equation*}
\begin{aligned}
    \sum_{m = 0}^{q_i} \tilde{w}_i(m,\mathcal{J}) 
    &= q_i \, 2^{i} + \tilde{w}_i(q_i,\mathcal{J})\\
    &\in 
    \begin{cases}
      \mleft\{q_i \, 2^{i} + s_i,\, q_i \, 2^{i} + r_i - s_i\mright\} & \text{if } q_i > 0,\\
      \mleft\{q_i \, 2^{i} + s_i\mright\} & \text{if } q_i = 0.
    \end{cases}
\end{aligned}
\end{equation*}  
Subsequently, since $q_i > 0 \iff 2^{i+1} \leq b$ and
\begin{equation*}
    w_i(b) = q_i \, 2^{i} + s_i = b - q_i \, 2^{i} - r_i + s_i,
\end{equation*}
the codeword weights of coset $i$ are $\mathcal{W}_i(\boldsymbol{B}_{k,a}) = \mathcal{V}_i(b)$.
\hfill\IEEEQED

\subsection{Proof of Theorem~\ref{thm:anticode weight dist}}\label{app:thm:anticode weight dist}
We adopt the notations from the proof of Lemma~\ref{lem:coset weights}.
For indices $0 \le i < k$, each coset contains $2^{k-i-1}$ codewords, while the coset $i = k$ contains only the all-zero codeword.
It follows from Lemma~\ref{lem:coset weights} that coset $i$ has codeword weights $\mathcal{W}_i(\boldsymbol{B}_{k,a})$,
and the number of distinct weights is
\begin{equation*}
    \mleft\lvert\mathcal{W}_i(\boldsymbol{B}_{k,a})\mright\rvert
    = \begin{cases}
        2 & \text{if } q_i > 0 \text{ and } r_i > 0,\\
        1 & \text{otherwise.}
    \end{cases}
\end{equation*}
It suffices to consider the case where $q_i > 0$ and $r_i > 0$, since otherwise all codewords have the same weight $w_i(b)$.
It follows that $\boldsymbol{g}_{j,i+1}(q_i) \in \{\boldsymbol{0}, \boldsymbol{1}\}$ for $j > i$, and $\tilde{w}_i(q_i,\varnothing) \neq 2^i$.
Consequently, adding the row $\boldsymbol{f}_j = (\boldsymbol{b}_{b-r_i,0}\mid \boldsymbol{1})$ to a codeword, with $j = \lb(b)-1$, changes its weight.
Thus, the row~$\boldsymbol{f}_j$ pairs every message
\begin{equation*}
    \mathcal{J}_1 \in \mathcal{I}_1 = \bigl\{\mathcal{J} \subseteq (i,k) \;\big\vert\; \tilde{w}_i(q_i,\mathcal{J}) = s_i\bigr\}
\end{equation*}    
with a unique message
\begin{equation*}
    \mathcal{J}_2 = (\mathcal{J}_1 \setminus \{j\}) \cup (\{j\} \setminus \mathcal{J}_1) \in \mathcal{I}_2 = \{\mathcal{J} \subseteq (i,k) \} \setminus \mathcal{I}_1.
\end{equation*}        
Therefore, both codeword weights occur equally often as 
\begin{equation*}
    A_{{w_i(b)}}^{(i)}(\boldsymbol{B}_{k,a}) 
    = \lvert\mathcal{I}_1\rvert 
    = A_{{b-w_i(b)}}^{(i)}(\boldsymbol{B}_{k,a})
    = \lvert\mathcal{I}_2\rvert 
    = 2^{k-i-2}.
\end{equation*}
\hfill\IEEEQED

\subsection{Proof of Corollary~\ref{cor:NPRS weight dist}}\label{app:cor:NPRS weight dist}
An \ac{NPRS} code with generator matrix $\boldsymbol{G}$ is obtained by puncturing the repeated simplex code $\mathcal{S}_k(\lceil n/M_k\rceil)$.
Since all non-zero codewords of $\mathcal{S}_k(\lceil n/M_k\rceil)$ have weight $d$, \eqref{eq:weight by anti} gives
\begin{equation*}
  \w(\boldsymbol{m}\boldsymbol{G}) =
  \begin{cases}
    0 & \text{if } \boldsymbol{m}=\boldsymbol{0},\\
    d-\w(\boldsymbol{m}\boldsymbol{B}_{k,a(n,k)}) & \text{otherwise},
  \end{cases}
\end{equation*}
for $\boldsymbol{m} \in \mathbb{F}_2^k$. Hence, each non-zero message that generates an \ac{NPRS} codeword of weight $w$, generates an anticode codeword of weight $d-w$. 
By Theorem~\ref{thm:anticode weight dist}, the latter are counted by the cosets $0 \leq i < k$, yielding the weight distribution.
\hfill\IEEEQED
\end{document}